\documentstyle[aps,preprint,epsf]{revtex}

\begin{document}
\draft \tightenlines

\def\be{\begin{equation}}
\def\ee{\end{equation}}
\def\bea{\begin{eqnarray}}
\def\eea{\end{eqnarray}}
\def\pd{\partial}
\def\a{\alpha}
\def\b{\beta}
\def\g{\gamma}
\def\d{\delta}
\def\m{\mu}
\def\n{\nu}
\def\t{\tau}
\def\l{\lambda}

\def\s{\sigma}
\def\e{\epsilon}
\def\scri{\mathcal{J}}
\def\cM{\mathcal{M}}
\def\tcM{\tilde{\mathcal{M}}}

\preprint{KUNSAN-TP-00-9}

\title{Brane inflation in tachyonic and non-tachyonic type 0B string theories}

\author{ Jin Young Kim\footnote{Electronic address:
jykim@ks.kunsan.ac.kr}}
\address{Department of Physics, Kunsan National University,
Kunsan, Chonbuk 573-701, Korea}

\maketitle

\begin{abstract}

 We consider the motion of the brane universe moving in a
background bulk space of tachyonic and non-tachyonic type 0B
string theory. The effective densities are calculated for both
cases and they show different power law behavior. Brane inflation
for non-tachyonic type 0B background has the same power law
behavior as that for type IIB background. The brane inflation
under tachyonic background is less divergent than the one without
tachyon. The role of tachyonic field in brane inflation scenario
is different from that of the ordinary matter field.

\end{abstract}

\pacs{PACS number(s): 11.10.Kk, 11.25.-w, 98.80.Cq }

\newpage

\section{Introduction}

Recently the old idea that our universe might be embedded in a
higher dimensional space \cite{RT} has attracted much interest.
The most attractive scenario is the so-called `Randall Sundrum
(RS) brane world' \cite{RS}. In this scenario our observed
universe is embedded in a five-dimensional bulk, in which the
background metric is curved along the extra dimension due to the
negative bulk cosmological constant. Under this framework
everything is confined to live on the brane except for gravity.
These models have been studied extensively because they might
provide the key to the gauge hierarchy problem and cosmological
constant problem \cite{KVK}.

It seems natural that one can generalize these models within a
well defined framework such as string theory or M-theory. Many
attempts have been made to apply this idea to string theory in the
context with D-branes \cite{Pol}. One of the important issues of
the brane world scenario is the cosmological evolution of the
early universe. Many cosmological models regarding this have been
suggested. These models can be classified into two categories. The
first is that the domain walls (branes) are static solution of the
underlying theory and the cosmological evolution of our universe
is due the time evolution of energy density on the domain wall
\cite{Tye}. The second is that the cosmological evolution of our
universe is due to the motion of our brane-world in the background
of gravitational field of the bulk \cite{CR,Kiritsis}. One of the
interesting among the second category is the the mirage cosmology
of Kehagias and Kiritsis \cite{Kiritsis}. The idea is that the
motion of the brane in ambient space induces cosmological
expansion (or contraction) on our universe.

There have been studies on how the presence of the various matter
fields on the supersymmetric background geometry affects the
cosmological evolution of the brane universe in the context of
type II theory \cite{Kiritsis,jykim}.
 The brane inflation for the case of non-supersymmetric string
 background has been studied with the type 0B theory in \cite{PP}.
 However the analysis is incomplete in the sense that their study
 is restricted to constant dilaton and tachyon field. The running
 of the tachyon is the key ingredient of the tachyonic type 0B
 theory \cite{KleTse,Minahan}. In this paper we will study the motion of a
 three-brane in the background of type 0B theory.
 We will consider
 both the tachyonic and non-tachyonic type 0B theories and see
 how the tachyonic field affects the brane inflation.

The organization of the paper is as follows. In Sec. II we will
extend the formalism of brane geodesic in Ref. \cite{Kiritsis} to
the case when tachyon field is present and set up some
preliminaries for our calculation. In Sec. III we consider the
type 0B background geometries. We will consider the tachyonic 0B
theory as well as the non-tachyonic 0B theory. In Sec. IV, using
the background solutions of Sec. III, we study the brane inflation
under these backgrounds. Finally in Sec. V, we discuss and
conclude our results.

\section{Brane geodesics }
In Ref. \cite{Kiritsis} it is shown that the motion of the brane
in ambient space induces cosmological expansion (or contraction)
on our universe simulating various kinds of matter or a
cosmological constant. In this section, we extend the formalism to
the case when there is tachyon field.

 We consider a probe brane
moving in a generic static spherically symmetric background. We
ignore its back reaction to the ambient space. As the brane moves
in a geodesic, the induced world-volume mertic becomes a function
of time. The cosmological evolution is possible from the brane
resident point of view. We will focus on a D3-brane case. For this
purpose we parametrize the metric of a D3-brane as
 \be
 ds^2_{10} = g_{00} (r) dt^2 + g(r) (d \vec x)^2 + g_{rr} (r) dr^2
 + g_S(r) d\Omega^2_5 ,
 \label{ds10gen}
 \ee
and we have dilaton $\phi$, tachyon $T(r)$ as well as RR
(Ramond-Ramond) background $C(r) = C_{0123} (r)$. In tachyonic
type 0B theory the D-brane effective action is determined not only
by the dilaton alone, but also by tachyon field. Due to the
existence of a tachyon tadpole on the D-brane, the Born-Infeld
(BI) part of the effective action of an electric D3-brane,
ignoring the fermions, is given by \cite{KleTse,garousi}
 \be
 S  =  T_3 \int d^4 \xi k(T) e^{- \phi} \sqrt{-{\rm det}( {\hat
G}_{\alpha\beta} + (2 \pi \alpha^\prime) F_{\alpha\beta} -
B_{\alpha\beta} ) }
  +  T_3 \int d^4 \xi {\hat C}_4 + \cdots ,
 \label{DBI}
 \ee
 where the induced metric on the brane is
 \be
 \hat G_{\alpha\beta} = G_{\mu\nu}
 { {\partial x^\mu} \over {\partial \xi^\alpha} }
 { {\partial x^\nu} \over {\partial \xi^\beta} }
 \ee
with similar expressions for other fields and
 \be
 k(T) =  1 + {1 \over 4} T + O(T^2) .
 \label{kt}
 \ee
 The reparametrization invariance and T-duality imply that the
 function $k(T)$ multiplies
 $\sqrt{-{\rm det}(\cdots ) }$ in the BI part of the
 action. The coefficient 1/4 of the tachyon tadpole was found in
 \cite{KleTse}.

Generally the motion of a probe D3-brane can have a nonzero
angular momentum in the transverse directions. We can write the
relevant part of the Lagrangian, in the static gauge $x^\alpha =
\xi^\alpha ~ ( \alpha = 0,1,2,3)$, as
 \be
 L = \sqrt{ A(r) - B(r) {\dot r}^2 - D(r) h_{ij} {\dot \varphi}^i
 {\dot \varphi}^j } - C(r),
 \ee
where $ h_{ij} \varphi^i \varphi^j $ is the line element on the
unit five sphere ($i,j=5, \dots, 9$),
 \be
 A(r)=g^3(r)|g_{00}(r)|e^{-2\phi} k^2(T) , ~
 B(r)=g^3(r)g_{rr}(r)e^{-2\phi} k^2(T) , ~
 D(r)=g^3(r)g_S(r)e^{-2\phi} k^2(T) ,
 \label{ABD}
 \ee
and $C(r)$ is the RR background. The momenta of the system are
given by
 \bea
 && p_r = -{B(r) \dot r \over
 \sqrt{A(r)-B(r)\dot r^2 - D(r) h_{ij} {\dot \varphi}^i
 {\dot \varphi}^j } } ,
    \nonumber \\
 && p_i  = -{D(r) h_{ij} \dot \varphi^j \over
 \sqrt{A(r)-B(r)\dot r^2 - D(r) h_{ij} {\dot \varphi}^i
 {\dot \varphi}^j } } .
 \eea
Calculating the Hamiltonian and demanding the conservation of
energy, we have
 \be
 H = C(r) - { A(r) \over \sqrt{A(r)-B(r)\dot r ^2-D(r)h_{ij} {\dot
 \varphi}^i {\dot \varphi}^j} } = - E ,
 \label{Hamil}
 \ee
 where $E$ is the total energy of the brane.
 Also from the conservation of the total angular momentum $h^{ij}p_ip_j=\ell^2$,
 we have
 \be
 h_{ij} {\dot \varphi}^i {\dot \varphi}^j = {\ell^2[A(r)-B(r)\dot
 r^2] \over  D(r)[D(r)+\ell^2]} .
 \label{lcons}
 \ee
Substituting Eq. (\ref{lcons}) into Eq. (\ref{Hamil}) and solving
with respect to $\dot r^2$, we have the equation for the radial
variable
 \be
 {\dot r}^2 = {A\over B} \left\{
 1-{A\over (C+E)^2}{D+\ell^2 \over D}
 \right\} .
 \label{rsol}
 \ee
 Plugging Eq. (\ref{rsol}) back into Eq. (\ref{lcons}), we
 have the equation for the angular variable
 \be
 h_{ij} {\dot \varphi}^i {\dot \varphi}^j
 = {A^2\ell^2\over D^2(C+E)^2} .
 \label{lsol}
 \ee

The induced four-dimensional metric on the three-brane universe is
 \be
 d \hat s^2_{4d} = (g_{00}+g_{rr} \dot r^2
 + g_S h_{ij} {\dot \varphi}^i {\dot \varphi}^j )
 dt^2+g(d\vec x)^2 .
 \label{4dmet}
 \ee
Using Eqs. (\ref{rsol}) and (\ref{lsol}), this reduces to
 \be
 d\hat s^2_{4d}=-{g^2_{00}g^3 e^{-2\phi}k^2(T) \over (C+E)^2}dt^2
 + g(d\vec x)^2 \equiv -d \eta^2 +g(r(\eta))(d\vec x)^2 ,
 \ee
 where we defined, for the standard form of
a flat expanding universe, the cosmic time $\eta$ as
 \be
 d\eta={|g_{00}|g^{3/2}e^{-\phi}|k(T)| \over |C+E|}dt .
 \ee
If we define the scale factor as $a^2=g$, we can calculate, from
the analogue of the four-dimensional Friedman equation, the Hubble
constant $H={\dot a / a}$
 \be
 \left({\dot a \over a}\right)^2
 ={(C+E)^2g_Se^{2\phi}k^{-2}(T)
 -|g_{00}|[g_Sg^3 + \ell^2e^{2\phi} k^{-2}(T)]
 \over 4|g_{00}|g_{rr}g_Sg^3 }
 \left({g'\over g}\right)^2 ,
 \label{hub}
 \ee
 where the dot denotes the derivative
with respect to cosmic time and the prime denotes the derivative
with respect to $r$. The right hand side of Eq. (\ref{hub}) can be
interpreted as the effective matter density on the probe brane
 \be
 {8\pi \over 3}\rho_{\rm eff}
 ={(C+E)^2g_Se^{2\phi}k^{-2}(T)
    -|g_{00}|[g_Sg^3 + \ell^2e^{2\phi} k^{-2}(T)]
 \over 4|g_{00}|g_{rr}g_Sg^3 }
 \left({g'\over g}\right)^2 .
 \label{hubble}
 \ee
We also have
 \bea {\ddot a\over a}
 &=& \left(1+{g\over g'}{\partial\over \partial r}\right)
 {(C+E)^2g_Se^{2\phi} k^{-2}(T)
    -|g_{00}|[g_Sg^3+ \ell^2e^{2\phi} k^{-2}(T)]
   \over 4|g_{00}|g_{rr}g_Sg^3}
  \left({g'\over g}\right)^2 \nonumber \\
  &=& \left( 1+{1 \over 2} a {\partial\over \partial a} \right)
  {8\pi \over 3}\rho_{\rm eff} .
 \eea
Equating the above to
 $-(4\pi / 3)(\rho_{\rm eff} + 3p_{\rm eff})$,
 we can find the effective pressure
 \be
 p_{\rm eff} = - \rho_{\rm eff}
 - {1 \over 3}  a {\partial\over \partial a} \rho_{\rm eff}.
 \ee
The apparent scalar curvature of the four-dimensional universe is
 \be
 R_{4d}=6 \left\{ {\ddot a\over a}
 +  \left({\dot a \over a}\right)^2 \right\}
 = 8\pi\left(4+a\partial_a\right)\rho_{\rm eff} .
 \label{4dcurv}
 \ee

\section{The type 0B background solution}

\subsection{Tachyonic 0B background solution}

The tachyonic type 0B model has a closed string tachyon, no
fermions and a doubled set of R-R fields, and thus a doubled set
of D-branes \cite{BG97}. With the doubling of R-R fields, the
self-dual constraint on the five-form field is relaxed and one can
have D3 branes that are electric instead of dyonic. We consider
the low energy world volume action of $N$ coincident electric D3
branes. We start from the action for the tachyonic type 0B theory
\cite{KleTse,kim196}
 \bea
 S_0=  - { 1 \over 2 \kappa^2_{10}} \int
 d^{10} x \sqrt{ -g} \bigg[ R &-& {1 \over 2} \nabla_n \Phi
     \nabla^n \Phi
     - {1 \over 4}( \nabla_n T \nabla^n T + m^2 e^{(1 / 2) \Phi} T^2)
     \nonumber \\
    &-& { 1 \over 4 \cdot 5!} f(T) F_{n_1...n_5} F^{n_1...n_5} +
\cdots \bigg],
 \eea
 where $g_{mn}$ is the Einstein-frame metric,
$m^2 = - 2 / \alpha^\prime $, and the tachyon-R-R field coupling
function is
 \be
f(T)= 1+ T + {1 \over 2}T^2. \label{f(t)}
 \ee
The equations of
motion from this effective action are
 \be
 \nabla^2 \Phi ={1\over 8} m^2
e^{(1/2)\Phi} T^2,    \label{dileq}
 \ee
 \bea R_{mn} - {1 \over 2}
g_{mn} R &=& {1 \over 4} \nabla_m T \nabla_n T - {1\over 8} g_{mn}
[(\nabla T)^2 + m^2 e^{(1 / 2)\Phi} T^2] + {1\over 2} \nabla_m
\Phi \nabla_n \Phi  \nonumber \\
 &-& {1 \over 4} g_{mn} (\nabla
\Phi)^2 + {1 \over 4 \cdot 4!} f(T) (F_{m klpq} F_{n}^{~klpq}
 - { 1\over 10} g_{mn} F_{sklpq} F^{sklpq}),   \label{graeq}
 \eea
 \be
 (-\nabla^2 + m^2 e^{(1 / 2) \Phi} ) T +
  {1 \over 2 \cdot 5!} f'(T) F_{sklpq} F^{sklpq} = 0,  \label{taceq}
 \ee
 \be
 \nabla_m [ f(T) F^{mnkpq} ] =0.  \label{RReq}
 \ee

 If one parametrize the ten-dimensional (10D) Einstein-frame  metric as
 \be
ds^2_E = e^{(1 /2) \xi - 5 \eta} d \rho^2 + e^{-(1 / 2) \xi}
(-dt^2 + dx_i dx_i)
 + e^{(1 / 2)\xi - \eta} d\Omega^2_5,
 \ee
 where $\rho$ is the radial direction transverse to the three-brane
($i = 1,2,3$), then the radial effective action corresponding to
Eqs. (\ref{dileq}) - (\ref{RReq}) becomes
 \be
S=  \int d \rho \bigg[{1 \over 2} \dot \Phi^2
   + {1 \over 2} \dot \xi^2
- 5 \dot \eta^2 + {1 \over 4} \dot T^2
   - V(\Phi,\xi,\eta,T) \bigg],
 \ee
 \be
V = {1 \over 2 \alpha^\prime} T^2 e^{(1 / 2 )\Phi + (1 / 2) \xi -
5 \eta }
   + 20 e^{-4\eta} - Q^2 f^{-1} (T) e^{-2\xi}.
 \ee
 Here the constant $Q$ is the R-R charge and dot means the
derivative with respect to $\rho$. The resulting set of
variational equations
 \be
\ddot \Phi + {1\over 4 \alpha^\prime} T^2 e^{(1 / 2)\Phi + (1 / 2)
\xi - 5 \eta }=0, \label{Phiddot}
 \ee
 \be
\ddot \xi + {1\over 4 \alpha^\prime} T^2 e^{(1 / 2)\Phi + (1 / 2)
\xi - 5 \eta } + 2 Q^2 f^{-1} (T) e^{-2\xi}=0, \label{xiddot}
 \ee
 \be
\ddot \eta + 8 e^{-4\eta } + {1\over 4 \alpha^\prime} T^2 e^{(1 /
2)\Phi + (1 / 2)  \xi   - 5 \eta } =0, \label{etaddot}
 \ee
 \be
\ddot T + {2 \over \alpha^\prime} T e^{(1 / 2)\Phi + (1 / 2) \xi -
5 \eta }+ 2 Q^2 {f^\prime (T)\over f^2(T)} e^{-2\xi}=0,
\label{Tddot}
 \ee
 should be supplemented by the zero-energy
constraint
 \be
{1 \over 2} \dot \Phi^2 + {1 \over 2} \dot \xi^2 - 5 \dot \eta^2 +
{1 \over 4} \dot T^2
   +  V(\Phi,\xi,\eta,T) =0,  \label{zeroen}
 \ee
 which can be  used instead of one of the second-order
equations.

In the Einstein frame the dilaton decouples from the R-R terms
($|F_5|^2$) and the tachyon mass term plays the role of the source
term. It is easy to analyze the case where the $|F_5|^2$ is large
in Eq. (\ref{taceq}).  Assuming that the tachyon is localized near
the extremum of $f(T)$, i.e., $T=-1$, the asymptotic solution in
the UV($\rho \equiv e^{-y} \ll 1,~ y \gg 1$) region (near-horizon)
is given by \cite{KleTse}

 \be
 T = -1 + {8\over y} + { 4 \over y^2}( 39 \ln y - 20)
 + O \bigg ( { \ln^2 y \over y^3} \bigg ),  \label{tbsoluv}
 \ee
 \be
 \Phi = \ln (2^{15} Q^{-1}) - 2  \ln y  +
 {39 \over y} \ln y + O \bigg({ \ln y\over y^2 } \bigg), \label{dbsoluv}
 \ee
 \be
 \xi = \ln (2 Q) - y + {1\over y}
 + {1\over 2 y^2}( 39 \ln y - 104)
 + O \bigg( { \ln^2 y \over y^3} \bigg), \label{xbsoluv}
 \ee
 \be
 \eta = \ln 2 - {1 \over 2}y + {1\over y}
 + {1\over 2 y^2} (39 \ln y - 38)
 + O \bigg( { \ln^2 y \over y^3} \bigg ). \label{ebsoluv}
 \ee
 The 10D Einstein frame metric can be written as
 \bea
 ds^2_E &=& R^2_0 \bigg[ \bigg( 1 - {9\over 2y} - {351 \over
4y^2} \ln y + \cdots \bigg)   \bigg({1\over 4}dy \bigg)^2
\nonumber  \\
  &+& \bigg( 1 - { 1\over 2y} - { 39\over 4y^2} \ln y + \cdots \bigg )
  {e^{(1/ 2)y}\over 2 R^4_0} dx^\mu dx^\mu
 + \bigg( 1 - { 1\over 2y} - { 39\over 4y^2} \ln y + \cdots \bigg )
 d \Omega^2_5 \bigg],    \label{gbsoluv}
 \eea
 where $$ R^2_0 = 2^{-1/2} Q^{1/2} . $$
 Note that with $y = 4 \ln u$,
one can show that the metric is of the form of
${\rm AdS}_5 \times
{\rm S}^5$ at the leading order,
 \be
ds^2_E=  R^2_0 \bigg ( {du^2\over u^2} + {u^2\over 2 R^4_0} dx^\mu
dx^\mu + d \Omega^2_5  \bigg ).    \label{adsuv}
 \ee
 The corrections
cause the effective radius of ${\rm AdS}_5$ to become smaller than
that of ${\rm S}^5$. One can find the asymptotic freedom from the
large $u$ behavior of the leading effective gravity solution. It
is an important question whether it survives the full string
theoretic treatment. It has been argued that the solution does
survive due to the special structure related to the approximate
conformal invariance \cite{KleTse,Minahan}.  The crucial fact is
that the Einstein metric is asymptotic to ${\rm AdS}_5 \times {\rm
S}^5$. This geometry is
 conformal to flat space, so that the Weyl tensor vanishes in the
 large $u$ limit. Furthermore, both $\Phi$ and $T$ vary slowly for
 large $u$.

One interesting feature of the tachyon RG trajectory is that $T$
starts increasing from its critical value $T=-1$ from condition
$f^\prime (T) =0$. The precise form of the trajectory for finite
$\rho$ is not known analytically, but the qualitative feature of
the RG equation was analyzed by Klebanov and Tseytlin
\cite{KleTse}. Since $\Phi, ~\xi$ and $\eta$ have negative second
derivatives [see Eqs. (\ref{Phiddot}) -(\ref{etaddot})], each of
these fields may reach a maximum at some value of $\rho$. If, for
$\Phi$, this happens at finite $\rho$, then one reaches a peculiar
conclusion that the coupling is decreasing far in the infrared.
However this possibility was not realized. Instead, a different
possibility was realized: $\dot \Phi$ is positive for all $\rho$,
asymptotically vanishing as $\rho \rightarrow \infty$. They
succeeded in constructing the asymptotic form of such trajectory.
It is crucial that, as $\rho \rightarrow \infty$, $T$ approaches
zero so that, as in the UV region, $T^2 e^{(1/2) \Phi}$ becomes
small. For this reason the limiting Einstein-frame metric is again
${\rm AdS}_5 \times {\rm S}^5$. Thus the theory flows to a
conformally invariant point at infinite coupling. The shape of the
tachyon RG trajectory is that the tachyon starts at $T=-1$ at
$\rho=0$, and grows according to Eq. (\ref{tbsoluv}), then enters
an oscillating regime and finally relaxes to zero.

\subsection{non-tachyonic 0B background solution}

Recently there has been attempts to find a non-tachyonic
holographic description of non-supersymmetric gauge theory
\cite{armoni1,armoni2}. Here we will consider the background
geometry of non-tachyonic type 0B theory in Ref. \cite{armoni2}.
Non-tachyonic 0B theory is based on the framework of 0B strings
with a particular orientifold projection \cite{sagnotti} to
eliminate the tachyon from the bulk and to avoid the problems with
the doubling of the R-R sectors. The gauge theory from this
orientifold of type 0B model becomes conformal in the planar
limit, and several results can be copied from ${\cal N} = 4$. In
particular, the leading (planar) geometry is known to be of the
form $\rm{AdS_5} \times \rm{X^5} $. We consider a dual
gravitational description of the gauge theory in the spirit of the
AdS/CFT correspondence. The gravity description involves the
ten-dimensional metric, the dilaton and the R-R four-form
potential (with self-dual field strength) whose dymamic is
essentially encoded in the action \cite{armoni2}
 \be
 S= { 1 \over {(\alpha')}^4 } \int
d^{10} x \sqrt{ -g} \bigg[ e^{-2 \Phi}(R - 4 \nabla_n \Phi
\nabla^n \Phi)
      + {1 \over \alpha'} G e^{- \Phi}
      - { 1 \over 4} (\alpha')^4 | F_5 |^2 \bigg],
 \ee
 or in its equations of motion
 \be
e^{- 2\Phi}( R + 4 \nabla^2 \Phi - 4 (\nabla \Phi)^2 )
 + {1\over 2 \alpha'} G e^{- \Phi} = 0 , \label{ntdileq}
 \ee
 \be
 e^{- 2\Phi} ( R_{mn} + 2 \nabla_m \nabla_n \Phi )
 - {1 \over {4 \alpha'} } g_{mn} G e^{- \Phi}
 - {1 \over 4 \cdot 4!} ( F_{mklpq} F_{n}^{~klpq}
 - { 1\over 10} g_{mn} F_{sklpq} F^{sklpq}) = 0,   \label{ntgraeq}
 \ee
 \be
 \nabla_m ( F^{mnkpq} ) =0.  \label{ntRReq}
 \ee
The term $G e^{- \Phi} = G g_s^{-1}$ represents the dilaton
tadpole expected from half-genus in string perturbation theory.
 Although the
direct computation of this term is difficult, there are plausable
arguments in favor of its presence \cite{armoni2}. From now on we
set $\alpha' = 1$ for simplicity of calculation.

For the metric solution, we make the ansatz compatible with
four-dimensional Poincar\'e invariance
 \be
ds^2 = d \tau^2 + e^{2 \lambda (\tau)} (-dt^2 + dx_i dx_i)
 + e^{2 \nu (\tau)} d\Omega^2_5, \label{ntmetansatz}
 \ee
 where $\tau$ is the radial direction transverse to the three-brane
($i = 1,2,3$). The five-form field strength has $N - 1/2$ units of
flux induced by the $N$ D3 branes and the $O'3$ plane. We shall
ignore the contribution of the orientifold plane, manifestly
suppressed for large $N$. Inserting the above ansatz in Eqs.
(\ref{ntdileq}) and (\ref{ntgraeq}) and allowing for a $\tau$
dependence of the various fields, we have
 \be
 2 \ddot \Phi - 4 \ddot \lambda - 5 \ddot \nu
 - 4 {\dot \lambda}^2 - 5{\dot \nu}^2
 + {1 \over 2} N^2 e^{2 \phi - 10 \nu}
 - {1 \over 4} G e^\Phi = 0 ,
 \label{ntPhiddot}
 \ee
 \be
 \ddot \lambda
 - (2 \dot \Phi - 4 \dot \lambda - 5 \dot \nu) {\dot \lambda}
 - {1 \over 2} N^2 e^{2 \phi - 10 \nu}
 + {1 \over 4} G e^\Phi = 0 ,
 \label{ntlambdaddot}
 \ee
 \be
 \ddot \nu
 - (2 \dot \Phi - 4 \dot \lambda - 5 \dot \nu) {\dot \nu}
 - 4 e^{-2 \nu} + {1 \over 2} N^2 e^{2 \phi - 10 \nu}
 + {1 \over 4} G e^\Phi = 0 ,
 \label{ntnuddot}
 \ee
 \be
 4 {\dot \Phi}^2 + 12 {\dot \lambda}^2 + 20 {\dot \nu}^2
 - 16 {\dot \lambda} {\dot \Phi}
 - 20 {\dot \nu} {\dot \Phi}
 + 40 {\dot \lambda} {\dot \nu}
 - 20 e^{-2 \nu} - G e^\Phi
 + N^2 e^{2 \phi - 10 \nu} = 0 ,
 \label{ntzeroen}
 \ee
 where dot denotes the derivative with respect to $\tau$.
 Note that the $\tau \tau$ component (Eq. (\ref{ntzeroen}))
 of the Einstein's equation is
 not independent. This corresponds to the zero energy constraint
 for the associated mechanical system.  If we redefine
 $\Phi \to \Phi - \ln N$ to see the role of the different
 contributions of the low-energy action, the independent
 field equations are
 \be
 2 \ddot \Phi - 4 {\dot \Phi}^2
 + 8 \dot \Phi \dot \lambda + 10 \dot \Phi \dot \nu
 + 3 {G \over N} e^\Phi = 0 ,
 \label{ntPhiddot2}
 \ee
 \be
 \ddot \lambda
 - (2 \dot \Phi - 4 \dot \lambda - 5 \dot \nu) {\dot \lambda}
 - {1 \over 2} e^{2 \phi - 10 \nu}
 + {1 \over 4} {G \over N} e^\Phi = 0 ,
 \label{ntlambdaddot2}
 \ee
 \be
 \ddot \nu
 - (2 \dot \Phi - 4 \dot \lambda - 5 \dot \nu) {\dot \nu}
 - 4 e^{-2 \nu} + {1 \over 2} e^{2 \phi - 10 \nu}
 + {1 \over 4} {G \over N} e^\Phi = 0 .
 \label{ntnuddot2}
 \ee
One can understand the role of the dilaton tadpole from the above
equations: it represents a $1/N$ correction to the leading type
IIB supergravity  equations. In the large $N$ limit, we can solve
the equations by iterations. The solutions up to the
next-leading-order are
 \be
 \Phi = \varphi - {3 \over {8 \cdot 2^{3/8} } } {G \over N}
        e^{(5/4) \varphi} \tau ,  \label{ntdilsol}
 \ee
 \be
 \lambda = 2^{ 3/8} e^{-(1/4) \varphi} \tau
  +{3 \over 64} {G \over N} e^{ \varphi} \tau^2 ,
  \label{ntlamsol}
 \ee
 \be
 \nu = - {3 \over 8 } \ln 2 + { 1 \over 4} \varphi
 - {3 \over {32 \cdot 2^{3/8} } } {G \over N} e^{(5/4) \varphi}
 \tau ,  \label{ntnusol}
 \ee
and the metric tensor is
 \bea
 ds^2 &&= d \tau^2 +
  \exp \bigg [ 2 \cdot 2^{ 3/8} e^{-(1/4) \varphi} \tau
  +{3 \over 32} {G \over N} e^{ \varphi} \tau^2 \bigg ]
  ( -dt^2 + d{\vec x}^2 )    \nonumber  \\
 &&+ \exp \bigg [ - {3 \over 8 } \ln 2 + { 1 \over 2} \varphi
  - {3 \over {32 \cdot 2^{3/8} } } {G \over N} e^{(5/4) \varphi}
 \tau \bigg ] d \Omega_5^2 ,  \label{ntmetsol}
 \eea
 where $\varphi$ is a constant.
 Note that the introduction of a dilaton tadpole results into a
 running of dilaton and running radii for the  ${\rm AdS}_5$ and
 $\mathbf{R P}^5 ( = {\rm S}^5 / \mathbf{Z}_2 )$ \cite{armoni2},
where $\mathbf{Z}_2$ the non-tachyonic involution associated to
${\rm O}'3$.

 \section{Brane cosmology}

In this section we will consider the cosmology of probe D3-brane
when it is moving along a geodesic in the background of two type
0B solutions of the previous section.

\subsection{brane inflation under the tachyonic 0B background}

 The metric of D3-brane using the background
solution (\ref{gbsoluv}) can be written as
 \bea
 |g_{00}(y)| = g (y) &=&
 \bigg( 1 - { 1\over 2y} - { 39\over 4y^2} \ln y + \cdots \bigg )
  {e^{y/2} \over 2 R^2_0}     \nonumber   \\
 g_{yy} (y) &=& \bigg( 1 - {9\over 2y} - {351 \over
 4y^2} \ln y + \cdots \bigg) { R^2_0  \over 16}   \nonumber  \\
 g_S (y) &=&
   \bigg( 1 - { 1\over 2y} - { 39\over 4y^2} \ln y + \cdots \bigg )
    R^2_0  \label{tacuvmet}
 \eea
 To apply the formalism of Sec. II we also need to express RR field
 in terms of $y$. From the ansatz for the RR field
 \be
 C_{0123} = C(y), ~~ F_{0123y} = {d C (y) \over d y} ,
 \ee
Eq. (\ref{RReq}) can be integrated once
 \be
 {{d C(y)} \over dy} = C_1 g^{2}g^{-5/2}_S \sqrt{g_{yy}}
  f^{-1} (T)
 \ee
 where $C_1$ is a constant. One can easily check that $C_1 = 2 Q$
 is the right choice to match the parameters which we
 already used in Sec. III.
  Using the solution of the metric
 in (\ref{tacuvmet}), the
$RR$ field can be integrated with appropriate normalization
 \be
 C(y) = {1 \over Q} e^y
 \bigg( 1 - {2 \over y} + \cdots \bigg ) + Q_1 ,
 \label{Csoltac}
 \ee
where $Q_1$ is another integration constant. This constant can be
absorbed in the redefinition of energy $E' \equiv E + Q_1$

 Now we can calculate the effective density on the brane
 using Eqs. (\ref{tbsoluv}), (\ref{dbsoluv}),
 (\ref{tacuvmet}), and (\ref{Csoltac})
 \bea
 {8 \pi \over 3} \rho_{\rm eff}
  &=& \bigg( {8 \over Q} \bigg )^{1/2}
      \bigg ( 1 + {7 \over y} + \cdots \bigg )    \nonumber \\
 && \times \bigg[ \bigg\{ {1 \over Q}
 \bigg ( 1 - {2 \over y} + \cdots \bigg ) + { E' \over e^y }
   \bigg\}^2
 {2^{31} \over y^4} \bigg ( 1 - {1 \over 2y} + \cdots \bigg )
 k^{-2} (T)  \nonumber \\
 &&~~~ + \bigg ( 1 - {1 \over 2y} + \cdots \bigg )
 \bigg\{ \bigg ( 1 - {3 \over 2 y} + \cdots \bigg )
 + {\ell^2 \over Q}
 {2^{32} \over y^4} e^{-(3/2) y} k^{-2} (T)
  \bigg\}  \bigg].
 \eea
 If we substitute the scale $a^2  = g(y) $ and Eq. (\ref{kt}) to
 see the leading power behavior near the horizon region, we have
 \be
 {8 \pi \over 3} \rho_{\rm eff}
  = \bigg( {8 \over Q} \bigg )^{1/2}
   \bigg[ \bigg( {1 \over Q} + { E' \over {2Q a^4 } }
   \bigg )^2 {2^{27} \over 9(\ln a)^4}
    - \bigg\{ 1 + {\ell^2 \over {9 \sqrt{2} Q^{5/2}} }
 {2^{27} \over (\ln a)^4} a^{-6}
  \bigg\}  \bigg].
 \ee
 Near the black brane, one can see that
 $\rho_{\rm eff} \sim  1/ [a^8 (\ln a)^4] $.
 Without the logarithmic term, the cosmological expansion due
 to the brane motion is indistinguishable from that of type
 IIB background.

\subsection{brane inflation under non-tachyonic 0B background}

 We can repeat the same procedure for non-tachyonic case by setting $k(T)=1$.
 The dilaton and metric background solution for this case are
 \be
 \Phi = \varphi - {3 \over {8 \cdot 2^{3/8} } } {G \over N}
        e^{(5/4) \varphi} \tau ,  \label{ntdilsol2}
 \ee
 \bea
 |g_{00}(\tau)| = g (\tau) &=&
  \exp \bigg [ 2 \cdot 2^{ 3/8} e^{-(1/4) \varphi} \tau
  +{3 \over 32} {G \over N} e^{ \varphi} \tau^2 \bigg ] ,  \\
 g_{\tau\tau} &=& 1 ,   \\
 g_S (\tau) &=&  \exp \bigg [ - {3 \over 8 } \ln 2 + { 1 \over 2} \varphi
  - {3 \over {32 \cdot 2^{3/8} } } {G \over N} e^{(5/4) \varphi}
 \tau \bigg ] .
 \eea
 Also from the ansatz for the RR field
 \be
 C_{0123} = A(\tau), ~~ F_{0123\tau} = {d A (\tau) \over d \tau} ,
 \ee
 Eq. (\ref{ntRReq}) can be solved as
 \be
 A(\tau) = {N \over \sqrt{2}} e^{-\varphi} \exp
 \bigg[ 4 \cdot 2^{3/8} e^{-(1/4)\varphi} \tau + O(1/N) \bigg ] + A_1 ,
 \label{Csolnt}
 \ee
 where $A_1$ is an integration constant which can be absorbed in the
 redefinition of energy $E^\prime = E + A_1$. If we calculate the
 effective density on the brane, in the large $N$ limit, we have
  \be
 {8 \pi \over 3} \rho_{\rm eff}
  = 2^{3/4}  e^{-\varphi/2} \bigg [
  \bigg( {N \over \sqrt{2} }
  + { E' e^{2 \varphi} \over a^4 } \bigg )^2
  - ( 1 + \ell^2 2^{3/4}  e^{(3/2) \varphi} a^{-6} ) \bigg].
 \ee
 In the limit $a \to 0$, we have
 $ \rho_{\rm eff} \sim a^{-8} . $
 This power behavior agrees with the result of type II case
 \cite{Kiritsis}.

\section{Discussion}

 We considered the motion of a three-brane moving in a
background bulk space of type 0B string theory. Both the tachyonic
and non-tachyonic backgrounds are considered. The cosmological
pictures of two non-supersymmetric backgrounds are expected to be
different. This is because the Born-Infeld action, which determine
the motion of the brane, in the presence of the tachyonic field is
different from the one without it. We set up the the modified
formalism of brane inflation when there is tachyon field in the
background.

The effective density for non-tachyonic 0B background is
proportional to $a^{-8}$ as $a \to 0$. This is the same power law
behavior as that for type IIB background. The effective density
for tachyonic 0B background is $\rho_{\rm eff} \sim 1/[ a^8 (\ln
a)^4]$. Comparing the two results we conclude that the brane
inflation with tachyonic field in the background is less divergent
than the one without tachyon. At first sight, this seems
contradiction to our physical sense. If we add any matter field in
the ambient space the brane universe will inflate faster, compared
with the case without that field, due to the increased effective
density. This phenomenon was shown in type IIB theory by turning
on the electromagnetic energy on the brane \cite{Kiritsis} and by
turning on the axion field \cite{jykim}. It seems that this is
true for any ordinary matter field. However, this might not be
true when the field is tachyonic whose mass squared is negative.
The presence of the tachyonic field decreases the effective
density on the brane and slows down the inflation. So we conclude
that the role of tachyonic field in brane inflation scenario is
opposite to the ordinary matter field.

\section*{Acknowledgments}

We would like to thank E. Kiritsis for suggesting this study and
A. Armoni for useful discussion. This work was supported by the
Korean Physical Society.

\end{document}